\documentclass[cits]{PoS}
\pdfoutput=1
\usepackage{fixltx2e}

\title{Two-flavour lattice QCD correlation functions in the deconfinement transition region}

\ShortTitle{Two-flavour lattice QCD correlation functions in the deconfinement transition region}

\author{Bastian B. Brandt\\
        Institut f\"ur theoretische Physik, Universit\"at Regensburg, D-93040 Regensburg\\
        E-mail: \email{brandt@kph.uni-mainz.de}}

\author{\speaker{Anthony Francis} \\
        Institut f\"ur Kernphysik, Johannes Gutenberg-Universit\"at Mainz, D-55099 Mainz\\
        Helmholtz Institut Mainz, Johannes Gutenberg-Universit\"at Mainz, D-55099 Mainz \\
        E-mail: \email{francis@kph.uni-mainz.de}}

\author{Harvey B. Meyer and Hartmut Wittig \\
        PRISMA Cluster of Excellence, Johannes Gutenberg-Universit\"at Mainz, D-55099 Mainz \\
        Institut f\"ur Kernphysik, Johannes Gutenberg-Universit\"at Mainz, D-55099 Mainz \\
        Helmholtz Institut Mainz, Johannes Gutenberg-Universit\"at Mainz, D-55099 Mainz\\
        E-mail: \email{meyerh@kph.uni-mainz.de} and \email{wittig@kph.uni-mainz.de}}

\newcommand{\be}{\begin{equation}}
\newcommand{\ee}{\end{equation}}
\newcommand{\ba}{\begin{eqnarray}}
\newcommand{\ea}{\end{eqnarray}}
\newcommand{\la}{\label}

\newcommand{\<}{\langle}
\renewcommand{\>}{\rangle}

\abstract{We report on a lattice QCD calculation with two dynamical flavors of the
isovector vector correlator in the high-temperature phase.  We analyze
the correlator in terms of the associated spectral function by 
performing a fit for the difference of the thermal and vacuum spectral
functions, using also an exact sum rule that constrains this
difference. 
Additonally we carry out a direct fit for the thermal spectral
function, and obtain good agreement between the two analyses for
frequencies below the two-pion threshold. Under the assumption that
the spectral function is smooth in that region, we give an estimate of
the electrical conductivity.
}

\FullConference{Xth Quark Confinement and the Hadron Spectrum,\\
		October 8-12, 2012\\
		TUM Campus Garching, Munich, Germany}

\begin{document}

\section{Introduction}
The properties of strongly interacting matter under extreme conditions
are the subject of intensive experimental and theoretical
investigation. A comprehensive picture of a state of matter requires
not only the knowledge of equilibrium properties such as the equation
of state and static susceptibilities, but also an understanding of its
transport properties.

Here we report on recent progress in determining the Euclidean isovector 
vector correlation function using dynamical light quarks in the high temperature
phase of QCD. This enables the first analysis of its underlying spectral function with unquenched quarks. 
We are able to determine the gross features of the thermal spectral function by analyzing directly the difference
of the thermal and vacuum correlators. This difference can be further constrained using a recently derived sum rule \cite{Bernecker:2011gh}. In addition we employ an approach followed recently in a quenched study to determine the spectral function directly from the thermal correlator based on an appropriate Ansatz \cite{Ding:2010ga,Francis:2011bt,Ding:2013tmp}. 

Note, this proceedings article represents a shortened version of \cite{Brandt:2012jc} and we refer the reader to this publication for a more in depth discussion.

\section{Basic definitions and expectations}
%Our primary observables are the Euclidean vector current correlators,
%\ba
%G_{\mu\nu}(\tau,T) &=&   \int d^3x \; \< J_\mu(\tau,\vec x)\;J_\nu(0)^\dagger\>\,,
%\ea
%with $J_\mu(x)\equiv \frac{1}{\sqrt{2}}\left(\bar u(x)\gamma_\mu u(x) - \bar
%d(x)\gamma_\mu d(x)\right)$ the isospin current. 
%The quark number susceptibility is defined as
%\be
%\chi_s\equiv -\int d^4x\; \< J_0(x) J_0(0)\> = -\beta \int d^3x \;\< J_0(\tau,\vec x) J_0(0)\>.
%\label{eq:chis}
%\ee
%Due to charge conservation, the two correlators of interest are exactly related via
%\be
%G_{ii}(\tau,T)=\chi_s T + G_{\mu\mu}(\tau,T)~~.
%\label{eq:GiGv}
%\ee
%The Euclidean correlators have the spectral representation
%\be
%G_{\mu\nu}(\tau,T)  = \int_0^\infty \frac{d\omega}{2\pi} \; \rho_{\mu\nu}(\omega,T) \;
%\frac{\cosh[\omega(\beta/2-\tau)]}{\sinh(\omega\beta/2)}\;.
%\ee

Our primary observables are the Euclidean vector current correlators and their spectral representation:
\be
G_{\mu\nu}(\tau,T) =   \int d^3x \; \< J_\mu(\tau,\vec x)\;J_\nu(0)^\dagger\>\,=\int_0^\infty \frac{d\omega}{2\pi} \; \rho_{\mu\nu}(\omega,T) \;\frac{\cosh[\omega(\beta/2-\tau)]}{\sinh(\omega\beta/2)}\;
\ee
with $J_\mu(x)\equiv \frac{1}{\sqrt{2}}\left(\bar u(x)\gamma_\mu u(x) - \bar
d(x)\gamma_\mu d(x)\right)$ the isospin current.
For a given function $\rho(\omega,T)$, the reconstructed correlator is defined as
\be 
G^{\rm rec}(\tau,T;T') {\equiv} \int_0^\infty \frac{d\omega}{2\pi}\; \rho(\omega,T')\;
\frac{\cosh[\omega(\frac{\beta}{2}-\tau)]}{\sinh( \omega \beta/2)} \,.
\la{eq:Grec1-main}
\ee
It can be interpreted as the Euclidean correlator that would be
realized at temperature $T$ if the spectral function was unchanged between
temperature $T$ and $T'$.  For $T'=0$ it can be directly obtained from the
zero-temperature Euclidean correlator via~\cite{Meyer:2010ii}
\be
G^{\rm rec}(\tau,T) \equiv G^{\rm rec}(\tau,T;0) = 
\sum_{m\in\mathbb{ Z}} G(|\tau+m\beta|,T=0).
\la{eq:Grec2-main}
\ee
In the thermodynamic limit, the subtracted vector spectral function obeys a sum rule
(see~\cite{Bernecker:2011gh} sec.\ 3.2),
\be\la{eq:sr}
\int_{-\infty}^\infty \frac{d\omega}{\omega} \; \Delta\rho(\omega,T) = 0,
\qquad \qquad \Delta\rho(\omega,T) \equiv  \rho_{ii}(\omega,T)-\rho_{ii}(\omega,0).
\ee
The electrical conductivity of QCD, connected to the isospin diffusion constant $D$, is given by a Kubo formula in terms of the
low-frequency behavior of the spectral function, (where $C_{em} = \sum_{f=u,d} Q_f^2=5/9$)
\be
\sigma=C_{em}D\chi_s = \frac{C_{em}}{6}\lim_{\omega\to0} \frac{\rho_{ii}(\omega,T)}{\omega}.
\la{eq:kubo}
\ee

%%%%%%%%%%%%%%%%%%%%%%%%%%%%%%%%%%%%%%%%%%%%%%%%%%%%%%%%%%%%%%%%%%%%%%%%
%%%%%%%%%%%%%%%%%%%%%%%%%%%%%%%%%%%%%%%%%%%%%%%%%%%%%%%%%%%%%%%%%%%%%%%%
%%%%%%%%%%%%%%%%%%%%%%%%%%%%%%%%%%%%%%%%%%%%%%%%%%%%%%%%%%%%%%%%%%%%%%%%

\section{Lattice QCD data}

\begin{table}[t]
\centering
 % Give a unique label
% For LaTeX tables use
\begin{tabular}{|c|r|r|c|c|r|r|}
\hline
   &   $64^3\times 128$  & $64^3\times 16$  & Ref. &   &   $64^3\times 128$  & $64^3\times 16$ \\
\hline
$6/g_0^2$ & \multicolumn{2}{|c|}{ 5.50} & & $T \; [{\rm MeV}]$ &  & 253(4) \\ 
$\kappa$ & \multicolumn{2}{|c|}{ 0.13671 } & & $\chi_s/T^2$ &  &  0.871(1)  \\
$c_{\rm sw}$ & \multicolumn{2}{|c|}{ 1.7515} & & $A_1/T^3$ &  4.42(31) & \\
$Z_V$    & \multicolumn{2}{|c|}{  0.768(5)} & \cite{DellaMorte:2005rd} & $m_1/T$ &  3.33(5) & \\
$a[\textrm{fm}]$ & \multicolumn{2}{|c|}{ 0.0486(4)(5)} &   \cite{Fritzsch:2012wq} & $\kappa_0$    & 1.244(5)& \\
$m_\pi$[MeV]   &  \multicolumn{2}{|c|}{270}   & \cite{Fritzsch:2012wq} & $\Omega/T$ & 5.98(11) & \\
\hline
\hline
\end{tabular}
\caption{\it{% $N_{conf}[N_\tau=16]$ & 317 & \\
The left part of the table shows the common quantities characterizing 
 the zero-temperature and finite-temperature ensembles.  In the right part,
the fit parameters for the vacuum correlator in units of $T=253{\rm
  MeV}$ and the value of the (isospin) quark number susceptibility
$\chi_s/T^2$ are given.  For more details on the generation of the $N_\tau=128$ ensemble,
see \cite{Fritzsch:2012wq}.  The number of configurations generated
with $N_\tau=16$ is 317. }}
\label{tab:pars}      
\end{table}

All our numerical results were computed on dynamical gauge
configurations with two light, mass-degenerate
$\mathcal{O}(a)$-improved Wilson quark flavors. 
We calculated correlation functions using the same discretization and
masses as in the sea sector in two different ensembles. The first
corresponds to virtually zero-temperature on a $64^3\times 128$
lattice (labeled O7 in~\cite{Fritzsch:2012wq}) with a lattice spacing
of $a=0.0486(4)(5)$fm~\cite{Fritzsch:2012wq} and a pion mass of
$m_\pi=270$MeV, so that $m_\pi L = 4.2$.  Secondly we generated an
ensemble on a lattice of size $64^3\times 16$ with all bare parameters
identical to the zero-temperature ensemble.  In this way it is
straightforward to compare the correlation functions respectively in
the confined and deconfined phases of QCD.  Choosing $N_\tau=16$
yields a temperature of $T\simeq250$MeV. Based on preliminary results
on the pseudo-critical temperature $T_c$ of the crossover from the
hadronic to the high-temperature phase~\cite{Brandt:2012sk}, the
temperature can also be expressed as $T/T_c\approx 1.2$.

The vacuum correlator serves as a reference in this work.  To fix the
parameters of the lightest vector state in the finite volume of the
simulation, we fitted the vacuum correlation function to an
 Ansatz of the form\footnote{For details please see \cite{Brandt:2012jc}}
\be
G_{ii}(\tau,0)= A_1 e^{-m_1\tau} + \frac{3}{4\pi^2}\kappa_0\;\exp(-\Omega\tau)\;
\left( \Omega^2/\tau + 2\Omega/\tau^2 + 2/\tau^3\right).
\label{eqn:efmfit}
\ee

In addition we estimate the thermal (isovector) quark number
susceptibility $\chi_s/T^2$ from the time-time component of the vector
correlation function.  The parameters used in our
lattice setup, the `$\rho$-meson' parameters and the value of the
static susceptibility are summarized in Tab.~\ref{tab:pars}.

%%%%%%%%%%%%%%%%%%%%%%%%%%%%%%%%%%%%%%%%%%%%%%%%%%%%%%%%%%%%%%%%%%%%%%%%
%%%%%%%%%%%%%%%%%%%%%%%%%%%%%%%%%%%%%%%%%%%%%%%%%%%%%%%%%%%%%%%%%%%%%%%%
%%%%%%%%%%%%%%%%%%%%%%%%%%%%%%%%%%%%%%%%%%%%%%%%%%%%%%%%%%%%%%%%%%%%%%%%
\subsection{Thermal and vacuum correlators}
\label{sec:data}

\begin{figure}[t]
\centerline{
\includegraphics[width=.53\textwidth]{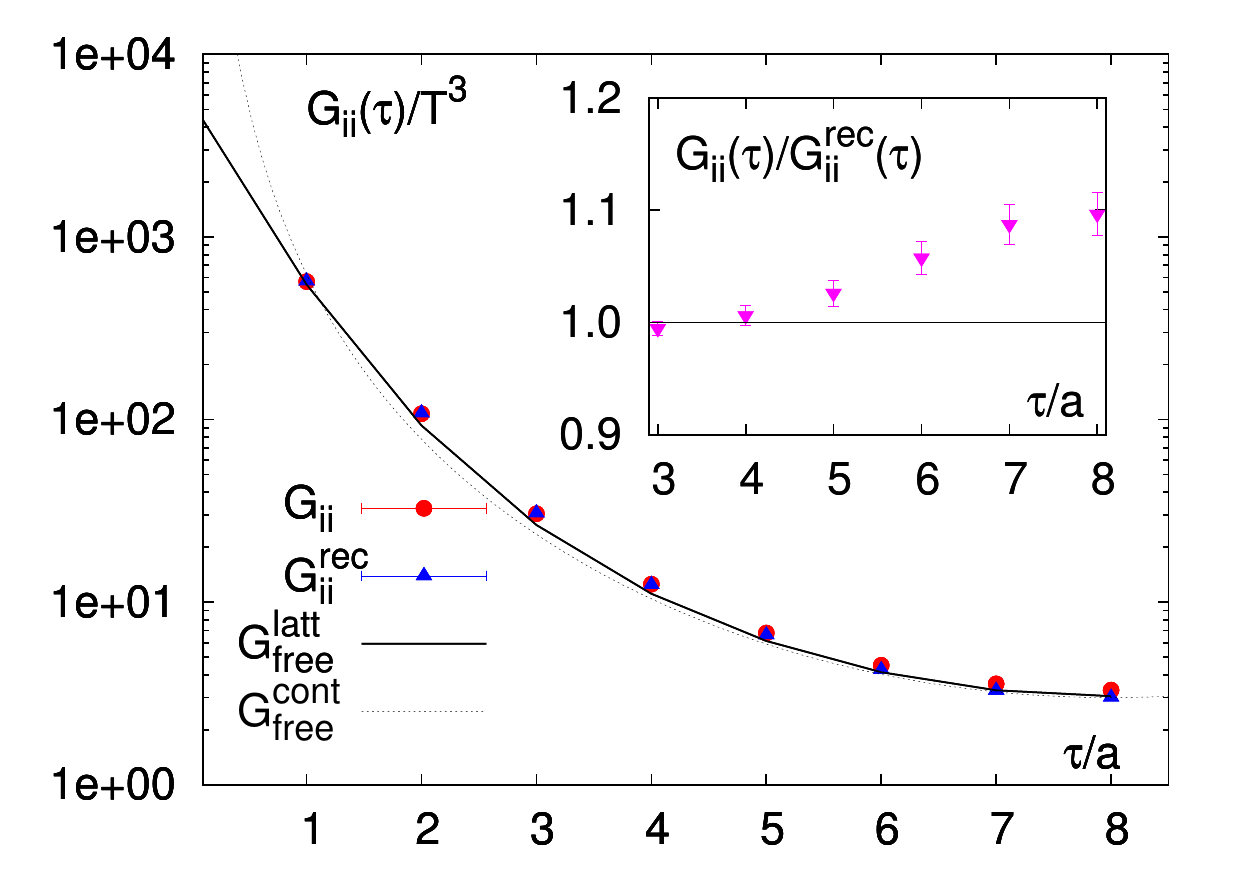}
\hspace*{-0.5cm}
\includegraphics[width=.53\textwidth]{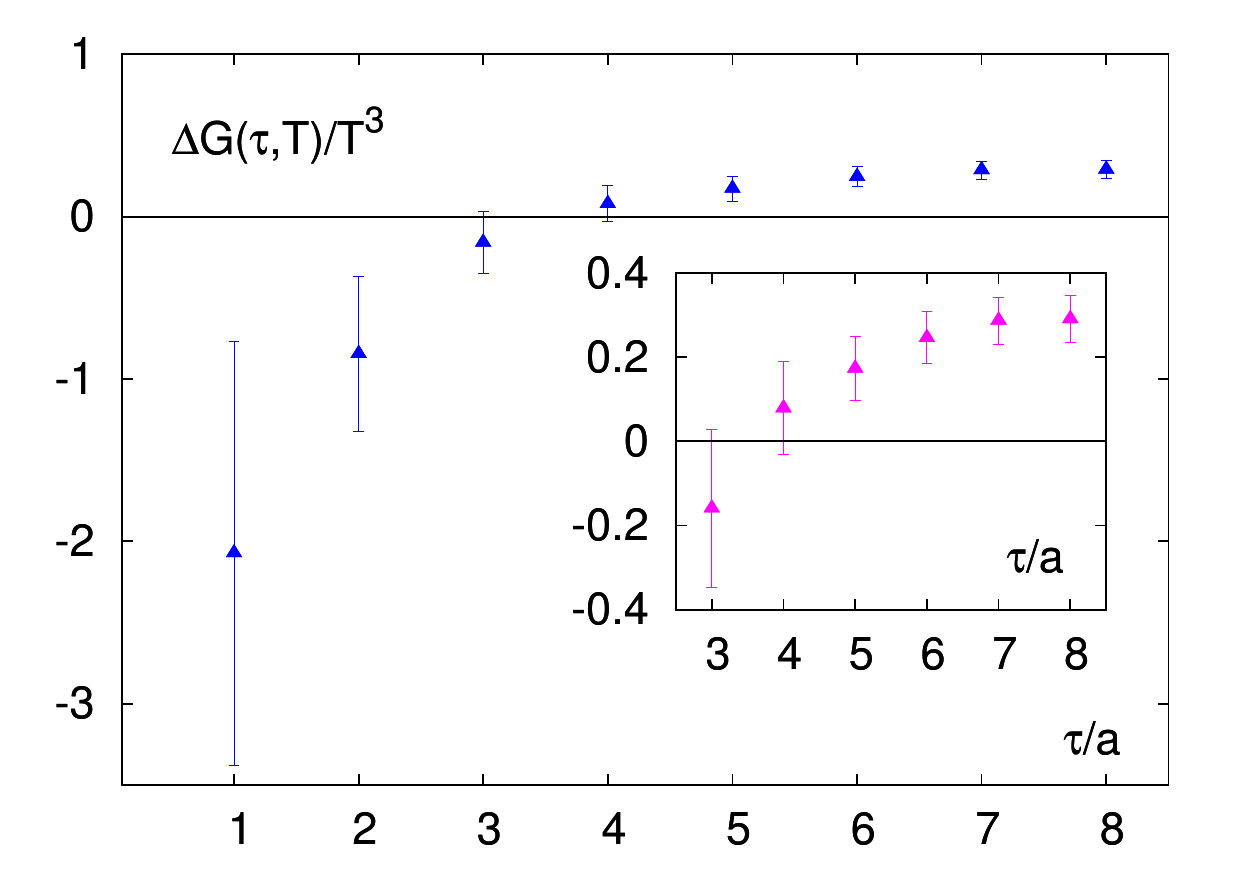}}
\caption{\it { Left: Thermal $G_{ii}(\tau)/T^3$ and reconstructed
    $G_{ii}^{rec}(\tau)/T^3$ vector correlators over Euclidean time
    separation $\tau$ compared to the free (continuum and lattice)
    cases. The reconstructed correlator was computed by applying 
    Eq.~\ref{eq:Grec2-main} to the data obtained from a lattice sized
    $N_\sigma=64$ and $N_\tau=128$ . The insertion shows the ratio
    $G_{ii}(\tau)/G_{ii}^{rec}(\tau)$. Right: Difference   $\Delta G(\tau,T)/T^3$  
  of the thermal vector correlator and the
  corresponding reconstructed correlator
  as a function of Euclidean time $\tau$. 
  The insertion shows $\Delta G(\tau,T)/T^3$ in the region $\tau/a \ge 3$. }}
\label{fig:Correlators}
\end{figure}

In Fig.~\ref{fig:Correlators}(left) we show the correlator $G_{ii}(\tau,T)$
computed at $T\simeq250$MeV together with the corresponding free
`continuum' and free `lattice discretized' correlation functions. In
addition we show the reconstructed correlator $G_{ii}^{rec}(\tau)$ as
obtained from Eq.~\ref{eq:Grec2-main}.
The reconstructed correlator lies somewhat lower than the thermal
correlator.  The insert in Fig.~\ref{fig:Correlators} displays the
ratio $G_{ii}(\tau)/G_{ii}^{rec}(\tau)$ in order to make their
relative $\tau$ dependence visible. For small Euclidean times
$\tau<\beta/4$ this ratio is unity, above it increases monotonically
until it levels off around the midpoint at about $10\%$ above unity.
A thermal modification of the spectral function has thus taken place
(recall that the spectral function underlying $G_{ii}^{rec}(\tau)$
contains the bound states of the confined theory).
In Fig.~\ref{fig:Correlators}(right) we show the difference 
\be
\Delta G(\tau,T) \equiv 
G_{ii}(\tau,T)-G_{ii}^{rec}(\tau,T)=\int_0^\infty \frac{d\omega}{2\pi} 
\Delta\rho(\omega,T)
\frac{\cosh[\omega(\beta/2-\tau)]}{\sinh(\omega\beta/2)}
\label{eq:GdiffGrec}
\ee
of the thermal and the reconstructed correlators. Given that
$\rho_{ii}(\omega,T)$ and $\rho_{ii}(\omega,T=0)$ have the same
$\sim\omega^2$ behavior, this means we are subtracting
non-perturbatively the ultraviolet tail of the spectral
function. Using this difference we are therefore able to probe
the change in the vector spectral function from the confined
to the deconfined phase for frequencies $\omega \lesssim {\rm O}(T)$.

For small times, the difference (\ref{eq:GdiffGrec}) turns out to be negative,
while it is positive for $\tau\geq\beta/4$. Note the
errors decrease with increasing Euclidean time throughout the
available range.  We show a more detailed view of the region $\tau/a\geq 3$
in the insert of Fig.~\ref{fig:Correlators}(right). Here the difference
still exhibits a mild increase and levels off near the midpoint.
The value it reaches at the midpoint is
$\Delta G(\tau=\beta/2,T)/T^3=0.291(55)$.

%%%%%%%%%%%%%%%%%%%%%%%%%%%%%%%%%%%%%%%%%%%%%%%%%%%%%%%%%%%%%%%%%%%%%%%%
%%%%%%%%%%%%%%%%%%%%%%%%%%%%%%%%%%%%%%%%%%%%%%%%%%%%%%%%%%%%%%%%%%%%%%%%
%%%%%%%%%%%%%%%%%%%%%%%%%%%%%%%%%%%%%%%%%%%%%%%%%%%%%%%%%%%%%%%%%%%%%%%%
\section{Analysis of lattice correlators in terms of spectral functions}

\subsection{Fit to the thermal part of the vector correlator}
\label{sec:GGrec}

We proceed to investigate the behavior of the thermal part of the
spectral function $\Delta\rho$ by fitting the difference of the
thermal and the reconstructed correlator, see
Eq.\ (\ref{eq:GdiffGrec}). As described in the previous
subsection, the fact that the data (displayed in
Fig.~\ref{fig:Correlators}(right)) is positive at long distances and negative
at short distances suggests that the thermal spectral weight exceeds
the vacuum spectral weight at low frequencies and falls short of it at
higher frequencies.

We thus parametrize $\Delta\rho$ using the following Ansatz for $\omega\geq 0$:
\ba \la{eq:ansatz2}
 \Delta\rho(\omega,T)&=&\rho_T(\omega,T)-\rho_B(\omega,T)+\Delta\rho_F(\omega,T), 
 \phantom{\frac{99}{17}}\\  \la{eq:ansatz3}
{\rho_B(\omega,T)} &=& \frac{2 c_B\, g_B\,\tanh(\omega/T)^3}{ 4(\omega-m_B)^2 + g_B^2 },
 \\   \la{eq:ansatz4}
 {\rho_{T,1}(\omega,T)} &=& \frac{4 c\,\omega}{ (\omega/g)^2 + 1 },\qquad \qquad ~~~~~\,
 {\rho_{T,2}(\omega,T)}=\frac{4c\,T\tanh(\omega/T)}{ (\omega/g)^2 + 1 },
\\    \la{eq:ansatz5}
\Delta\rho_F(\omega,T) &=& \rho_F(\omega,T) - \rho_F(\omega,0),\qquad 
{\rho_F(\omega,T)} = \frac{3}{2\pi}\kappa\,\omega^2 \tanh\left(\frac{\omega}{4T}\right)\;.
\ea
The bound state (B) and the transport peak (T) are represented by
Breit-Wigner forms. Even such a simple Ansatz requires three
parameters $(c_B,g_B,m_B)$ to determine the bound state peak, two
parameters $(c,g)$ for the transport peak and one $(\kappa)$ for the `perturbative'
contribution (F). We will therefore fix some of them using the vacuum
correlator.  In the following we set $m_B$ equal to $m_1$, given in
Tab.~\ref{tab:pars}, which we obtained from the exponential fit to the
vacuum correlator.  Note that the area under the bound state peak
$\int dw~\rho_B/\omega$ does not depend on the width $g_B$ in the
limit where it is small.  We therefore perform fits for
three fixed values of this parameter, and check the sensitivity of the
result.  We choose the values $g_B/T=0.1, 0.5$ and 1.0.

\begin{figure}[t]
\centerline{
\includegraphics[width=.53\textwidth]{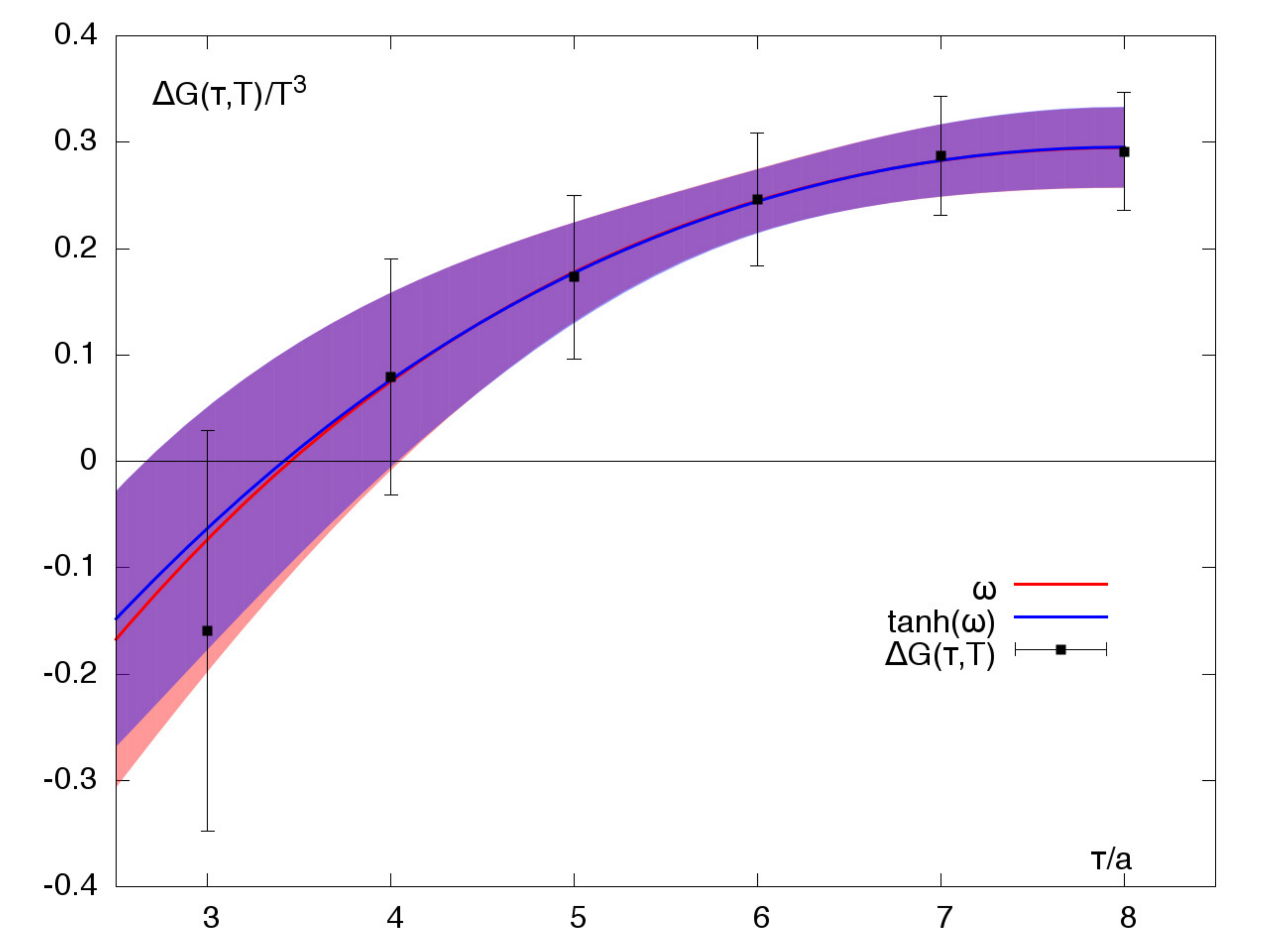}
\hspace*{-0.5cm}
\includegraphics[width=.53\textwidth]{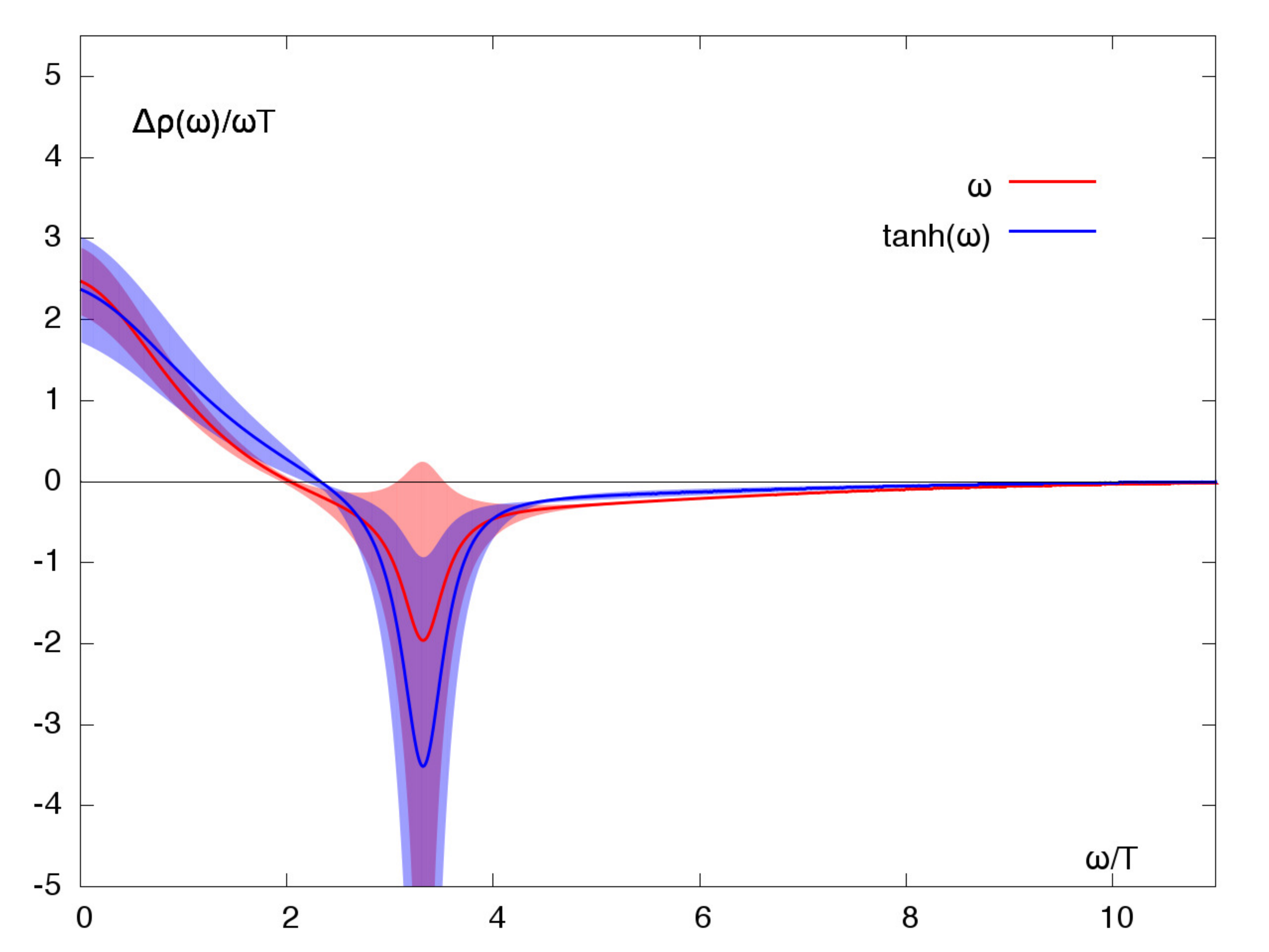}}
\caption{\it\small{Left: Fits to $\Delta G(\tau,T)/T^3\equiv
    [G_{ii}(\tau)-G_{ii}^{rec}(\tau)]/T^3$. The blue and red results
    differ by the form of transport peak in the Ansatz. The error
    bands are computed from the covariance matrix of the fit.  Right:
    The resulting spectral functions for both Ans\"atze.}}
\label{fig:GdiffGrec-CORSPF}
\end{figure}

The tail $\sim T/\omega$ of the Ansatz $\rho_{T,1}$ violates
the OPE prediction that $\Delta\rho\sim(T/\omega)^2$ at large
frequencies.  It has been argued in \cite{Burnier:2012ts} that this
might lead to an overestimate of the transport contribution. To avoid
this problem we introduce the Ansatz 2, where $\omega\rightarrow
T\tanh(\omega/T)$. This Ansatz possesses the correct asymptotic
behavior, as well as the expected linear behavior in $\omega$ at small
frequencies.
Finally, to complete the parametrization of $\Delta\rho(\omega)$, we
include a weak-coupling term
describing the subtraction of the large frequency parts of the thermal
and vacuum spectral functions. This contribution
$\rho_F(\omega,\kappa)\rightarrow 0$ vanishes exponentially as the
frequency increases.

In the next step we fit the combined Ans\"atze of
$\Delta\rho(\omega,c_B,g_B,m_B,c,g,\kappa)$ to the data, while at the
same time satisfying the sum rule of Eq.~\ref{eq:sr} to an accuracy of
$10^{-8}$.  We limit ourselves to fitting the region $5\leq\tau/a\leq8$
only, in order to minimize the influence of cut-off effects.  With $m_B$ determined by the vacuum
correlator, we set $g_B$ successively to the three different values
mentioned above and fixed  $\kappa$  around unity, and fitted
the parameters $c$, $g$ and $c_B$.  The errors and error bands shown
in the following have been computed using the covariance matrix of the
corresponding fit for fixed values of $g_B$ and $\kappa$.

The resulting correlators and spectral functions are displayed in
Fig.~\ref{fig:GdiffGrec-CORSPF} for $g_B/T=0.50$ and $\kappa=1.10$. In the
left panel of Fig.~\ref{fig:GdiffGrec-CORSPF} the data $\Delta
G(\tau,T)$ is compared to the fits using $\rho_{T,1}(\omega)$ and
$\rho_{T,2}(\omega)$ as transport contribution. We achieve a
quasi-perfect description of the data for $\tau/a\geq 4$.  The right
panel shows that both Ans\"atze exhibit a substantial spectral weight
around the origin and a negative contribution from the region of the
$\rho$ mass.

%%%%%%%%%%%%%%%%%%%%%%%%%%%%%%%%%%%%%%%%%%%%%%%%%%%%%%%%%%%%%%%%%%%%%%%%
%%%%%%%%%%%%%%%%%%%%%%%%%%%%%%%%%%%%%%%%%%%%%%%%%%%%%%%%%%%%%%%%%%%%%%%%
%%%%%%%%%%%%%%%%%%%%%%%%%%%%%%%%%%%%%%%%%%%%%%%%%%%%%%%%%%%%%%%%%%%%%%%%

\subsection{Weak-coupling inspired fit to the thermal vector\la{sec:fitG}}

In contrast to the previous section, here we study directly the
thermal vector correlator and its ratio of thermal moments
$R^{(2,0)}$, computed along the lines of \cite{Ding:2010ga}.
We perform a fit inspired by the weak coupling form of
the thermal spectral function,
\ba
\rho(\omega,T)&=&\rho_T(\omega,T)+\rho_F(\omega,T)~,
\ea 
where the form of the two contributions is defined in
Eq.\ (\ref{eq:ansatz4}) and (\ref{eq:ansatz5}).  At a given temperature this Ansatz is
characterized by three parameters $(c,g,\kappa)$.  We fit the full
Ansatz $\rho(\omega,c,g,\kappa)$ to the thermal correlator
$G_{ii}(\tau)$, while at the same time demanding that $R^{(2,0)}$ be
reproduced. In this analysis the three parameters $c, g$ and $\kappa$
are fitted, and the fit range is $5\leq\tau/a\leq 8$ as before.

\begin{figure}[t]
\centerline{
\includegraphics[width=.53\textwidth]{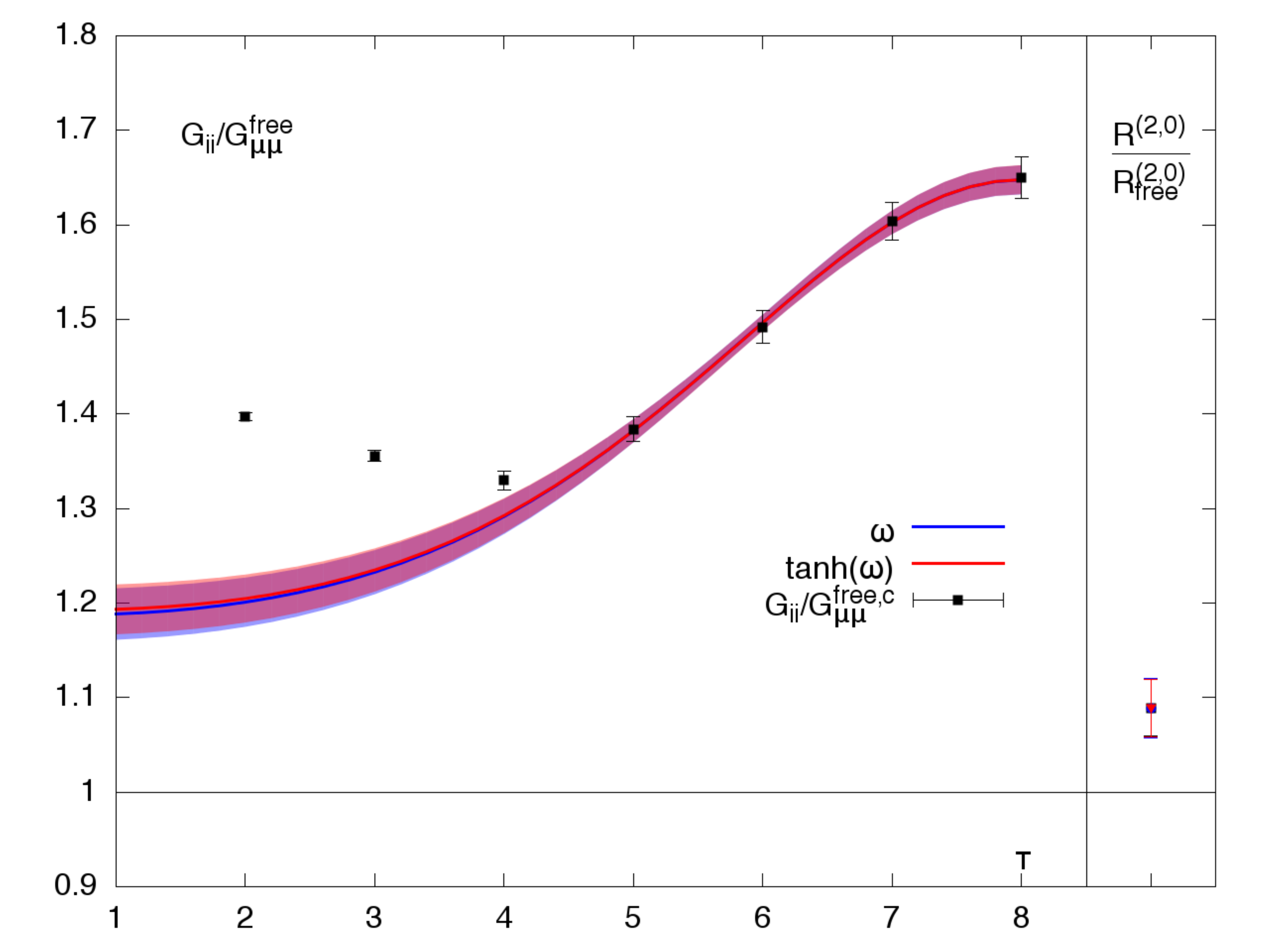}
\hspace*{-0.5cm}
\includegraphics[width=.53\textwidth]{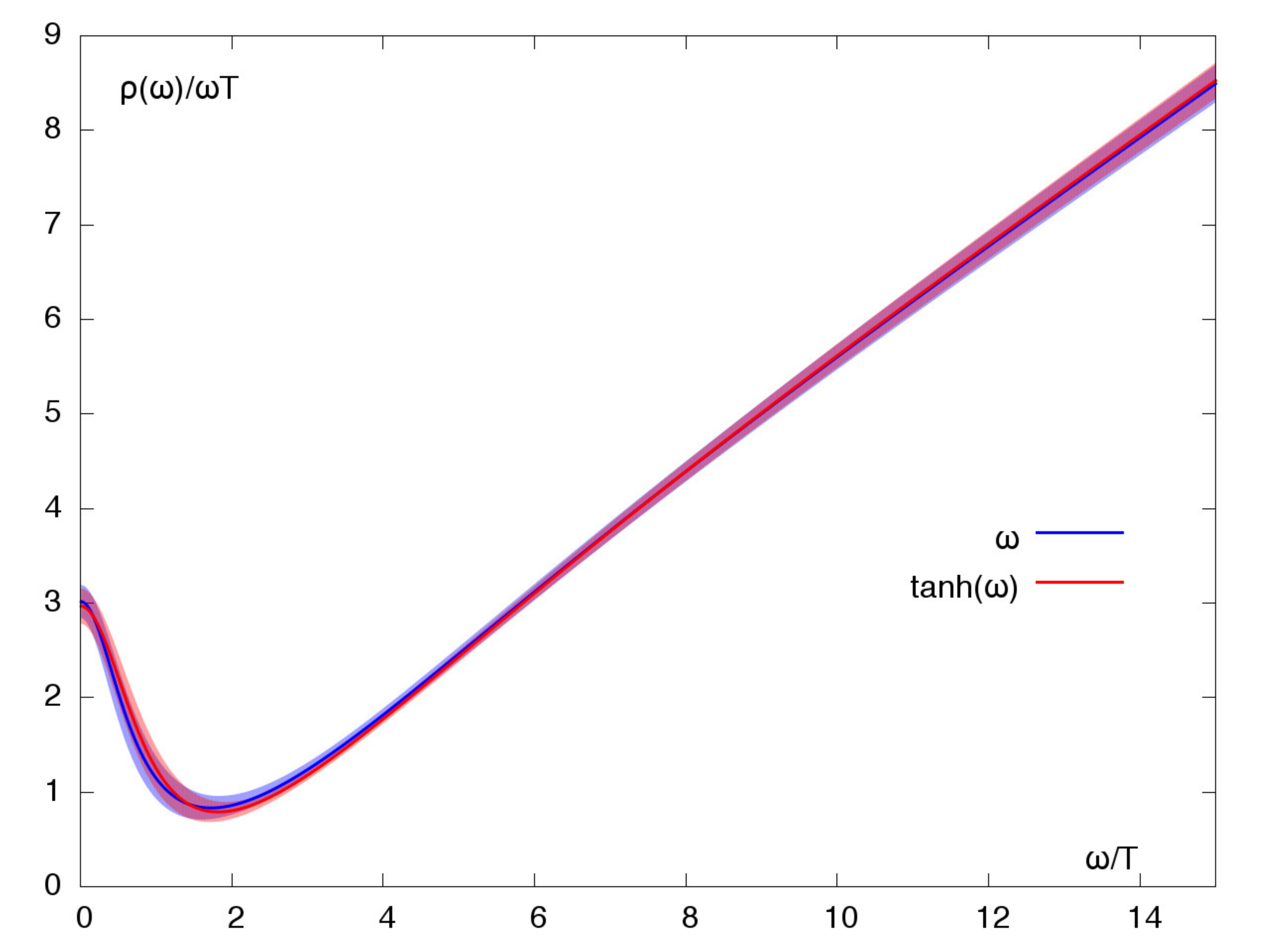}}
\caption{\it\small{Left panel: fit results for $G_{ii}(\tau)/G_{\mu\mu}^{free}(\tau)$ 
using both Ans\"atze for the transport peak. 
 The resulting ratio of thermal moments
 $R^{(2,0)}/R^{(2,0)_{free}}$ are displayed on the right side of the plot.
 Right panel: The corresponding spectral functions normalized by $\omega T$.}} 
\label{fig:Dileptonfit}
\end{figure}

The resulting correlators  and spectral functions are shown in
Fig.~\ref{fig:Dileptonfit}. The ratio $G_{ii}(\tau)/G_{\mu\mu}^{\rm
  free}(\tau)$ in the left panel of Fig.~\ref{fig:Dileptonfit} is well
described by both versions $\rho_{T,1}$ and $\rho_{T,2}$ of the
transport contribution for $\tau/a\geq5$, while also the ratio of
thermal moments (given on the far right of the plot) is reproduced.
For $\tau/a<5$ our Ansatz fails to reproduce these points, which we
suspect is partly due to cutoff effects.

On the right hand side of Fig.~\ref{fig:Dileptonfit} we show the
resulting spectral functions divided by $\omega T$.  Clearly both
Ans\"atze give very similar results that lie within errors of each
other. The thermal correlator is even less sensitive to the 
asymptotic behavior of the transport contribution in the Ansatz
than in the difference of correlators studied in Sec.~\ref{sec:GGrec}.

%%%%%%%%%%%%%%%%%%%%%%%%%%%%%%%%%%%%%%%%%%%%%%%%%%%%%%%%%%%%%%%%%%%%%%%%
%%%%%%%%%%%%%%%%%%%%%%%%%%%%%%%%%%%%%%%%%%%%%%%%%%%%%%%%%%%%%%%%%%%%%%%%
%%%%%%%%%%%%%%%%%%%%%%%%%%%%%%%%%%%%%%%%%%%%%%%%%%%%%%%%%%%%%%%%%%%%%%%%
\section{Discussion}

\begin{figure}[t]
\centering
\includegraphics[width=.63\textwidth]{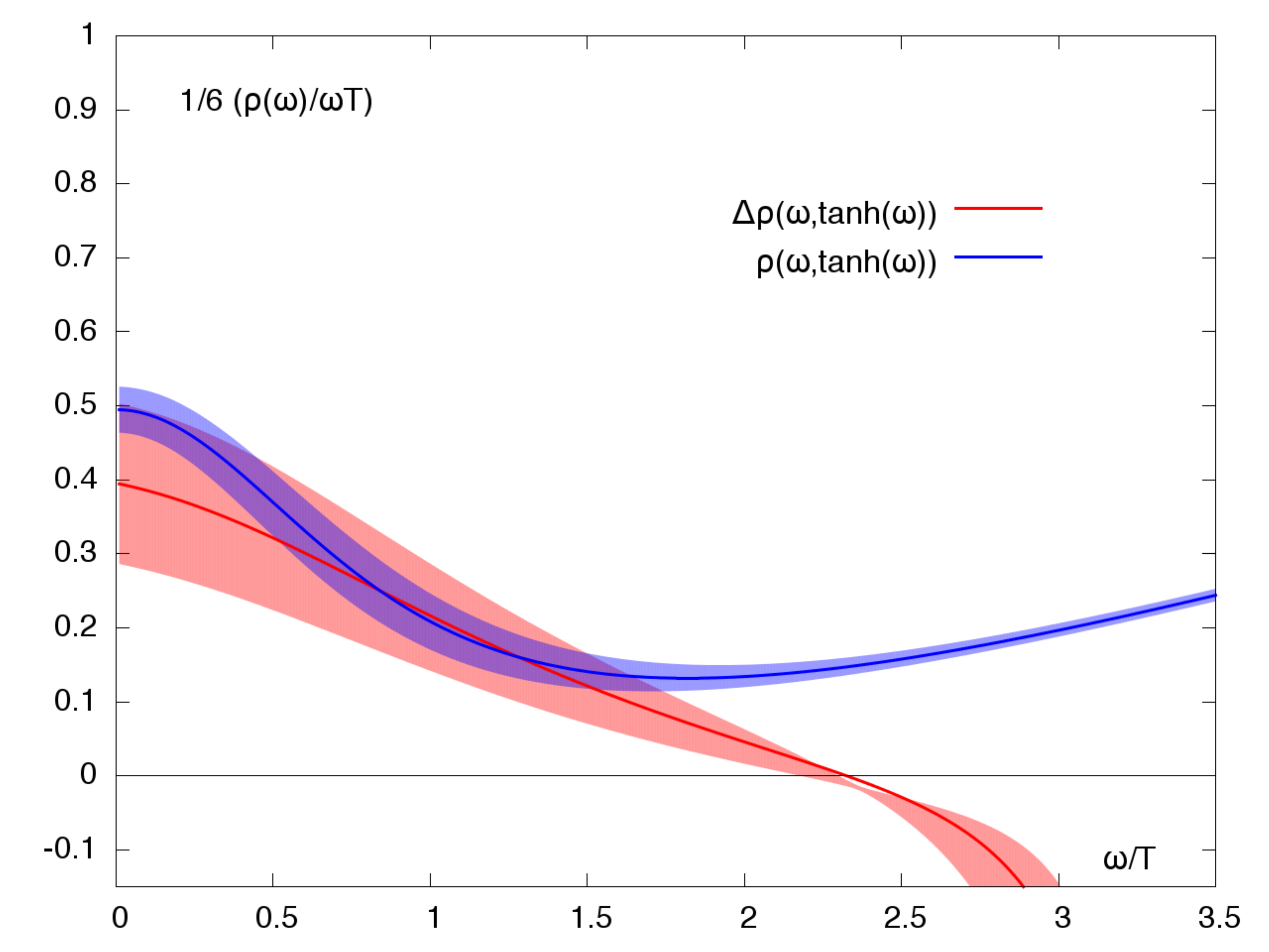}
\caption{\it\small{Comparison of the spectral functions obtained from analyzing
 (a) $\Delta G(\tau,T)$ and (b) $G_{ii}(\tau,T)$ 
using in both cases $\rho_{T,2}(\omega)\sim \tanh(\omega/T)$ in the low frequency region. 
All curves have been multiplied by a factor 1/6 and divided by $\omega T$, 
entailing that the intercept at $\omega=0$ yields an estimate of $\sigma/C_{em}T$.}}
\label{fig:Spectralfunctions}
\end{figure}

We now compare the results of the fits to $\Delta G(\tau,T)$ and
$G_{ii}(\tau,T)$. Since the vacuum spectral function vanishes below
$2m_\pi\approx 540{\rm MeV}$ (in infinite volume), $\rho_{ii}(\omega,T)$
and $\Delta\rho(\omega,T)$ should be equal for $\omega<2m_\pi\approx
2.1T$. We thus plot the spectral functions obtained from the two fits in
this frequency region, see Fig.~\ref{fig:Spectralfunctions}.  Here we
restrict ourselves to showing these results based on
$\rho_{T,2}(\omega,T)$, as their theoretical foundation is more sound
than those with $\rho_{T,1}(\omega,T)$.  All curves are multiplied by
a factor $1/6$, which means that the intercept at $\omega=0$ yields an
estimate of $\sigma/C_{em}T$ with $\sigma$ the electrical conductivity
of the quark gluon plasma.

The results obtained by fitting $G_{ii}(\tau,T)$ agree very well with
the central values obtained by fitting $\Delta G(\tau,T)$, whereby the fit to $\Delta G(\tau,T)$ using
the transport Ansatz $\rho_{T,2}(\omega)$ yields a slightly lower
intercept.  If we assume the spectral function to be as smooth around
the origin as Fig.~\ref{fig:Spectralfunctions} suggests, we obtain the
following estimate for the electrical conductivity of the quark gluon
plasma at $T\simeq 250$MeV,
\be \la{eq:sigma_res} 
\frac{\sigma}{C_{em}T}= 0.40(12),
\ee 
where $C_{em} = \sum_{f=u,d} Q_f^2$.
Although obtained under a strong assumption, it is interesting to
compare (\ref{eq:sigma_res}) to other lattice results obtained under
similar assumptions.  
The following comparison is made with quenched results, since 
to our knowledge there are no previous dynamical QCD studies.

Quenched calculations using staggered fermions and different methods for analyzing the spectral function obtained
$\sigma/T=7C_{em}$\cite{Gupta:2003zh} and $\sigma/T=0.4(1)C_{em}$
\cite{Aarts:2007wj} in the temperature region $1.5\leq T/T_c\leq3.0$. Also, a recent quenched study using
Wilson-Clover fermions in the continuum limit obtained
$0.33C_{em}\leq\sigma/T\leq 1C_{em}$ at $T\simeq1.45T_c$
\cite{Ding:2010ga} and $T\simeq1.1T_c$\cite{Ding:2013tmp}, with similar non-continuum results up to temperatures $T\simeq2.98T_c$\cite{Francis:2011bt}. Our results using dynamical Wilson-Clover fermions
at $N_\tau=16$ are thus completely compatible with the recent quenched
results.

\section*{Acknowledgments}
Computations were carried out on the JUGENE computer of the Gauss Centre
for Supercomputing located at Forschungszentrum J\"ulich, Germany, within NIC project HMZ21, and the dedicated QCD platform ``Wilson'' at the Institute for Nuclear Physics,
University of Mainz. This work was supported by the
\emph{Center for Computational Sciences in Mainz} and by the DFG grant
ME 3622/2-1 \emph{Static and dynamic properties of QCD at finite
  temperature}.

\bibliographystyle{JHEP}
\bibliography{viscobib.bib}

%\begin{thebibliography}{99}
%\bibitem{...} 
%....
%\end{thebibliography}
\end{document}